\documentclass{ws-ijmpa}
\usepackage{amsmath}

\begin{document}

\markboth{Stefan Sint}
{The Kaon $B$-parameter in quenched QCD}

%\catchline{}{}{}{}{}
%
%%%%%%%%%%%%%%%%%%%%%%%%%%%%%%%%%%%%%%%%%%%%%%%%%%%%%%%%%%%%%%%%%%%%

\title{The Kaon $B$-parameter in quenched QCD\footnote{based on work done 
in collaboration with Petros Dimopoulos, Jochen Heitger, Filippo Palombi, 
Carlos Pena and Anastassios Vladikas.}}

\author{\footnotesize STEFAN SINT (ALPHA-collaboration)}

\address{Universidad Aut\'onoma de Madrid, Instituto  de F\'{\i}sica 
Te\'orica,\\
Facultad de Ciencias C-XVI, \\
E-28049 Cantoblanco, Madrid, Spain\\}

\maketitle

%\pub{Received (Day Month Year)}{Revised (Day Month Year)}

\begin{abstract}
I report on a recent determination by the ALPHA
collaboration of the kaon $B$-parameter
using lattice QCD with Wilson type quarks.
An effort is made to control all systematic
errors except for the quenched approximation.
The preliminary result for the renormalization group invariant
parameter is $\hat{B}_K = 0.834(37)$, which translates to 
$B_K^{\overline{\rm MS}}(2\,{\rm GeV})=0.604(27)$ 
in the ${\overline{\rm MS}}$ scheme with anticommuting $\gamma_5$.
%\keywords{Keyword1; keyword2; keyword3.}
\end{abstract}

\section{Introduction}

The kaon $B$-parameter remains an important ingredient
in current analyses of CP violation in the Standard 
Model~\cite{Battaglia:2003in}. It is is defined in QCD with
up, down and strange quarks by the matrix element
of a four-quark operator between kaon states, 
\begin{eqnarray}
   \left\langle \bar{K}^0\vert O^{\Delta S=2}_{\rm (V-A)(V-A)}\vert K^0\right\rangle &=& \frac 83 F_K^2m_K^2 B_K,\nonumber\\
O^{\Delta S=2}_{\rm (V-A)(V-A)} &=& \sum_\mu\left[\bar{s}\gamma_\mu(1-\gamma_5)d\right]^2.
\end{eqnarray}
While this matrix element is well-defined, its relation
to the full amplitude in the Standard Model
relies on the hypothesis that the charm quark can 
be treated as a heavy particle.
While the quality of this approximation 
is difficult to assess, it is typically assumed to be valid
at the five percent level, which sets
the scale for the precision to be attained by
a lattice determination of $B_K$.

%that a 5 per cent uncertainty due to this approximation
%should be assumed, which therefore sets the scale for
%the precision to be attained in lattice simulations.

\section{$B_K$ and Wilson type quarks}

Despite recent progress with other fermion formulations,
lattice QCD with Wilson type quarks remains
attractive because it is computationally
cheap and does not suffer from mixing between flavour and spin
degrees of freedom (unlike staggered fermions).
On the other hand, all axial symmetries are explicitly
broken, which leads to mixing of operators with opposite chirality,
and the possible occurence of unphysical fermion 
zero modes in the quenched approximation.
In order to avoid the latter problem in typical  
quenched simulations, the masses of pseudoscalar
mesons are typically heavier than the physical kaon mass.
Concerning the calculation of $B_K$, the mixing problem
is particularly annoying. Decomposing the
operator $O_{\rm (V-A)(V-A)}=O_{\rm VV+AA}-O_{\rm VA+AV}$ 
the relevant piece for $K^0$-$\bar{K}^0$ mixing is the parity even
component, which renormalizes as follows~\cite{Bernard:1987pr}:
\begin{eqnarray}
  \left[O_{\rm VV+AA}\right]_{\rm R} &=&
   Z_{\rm VV+AA}\left[O_{\rm VV+AA}+\sum_{i=1}^4 z_i\ O_i^{d=6} \right].
\end{eqnarray}
While the mixing problem can be solved by imposing axial continuum
Ward identities~\cite{Aoki:1999gw}, 
it represents a major obstacle for precision results.
Note that the parity-odd operator component $O_{\rm VA+AV}$ 
does indeed renormalize multiplicatively~\cite{Bernard:1987pr}. 
This can be exploited for the computation of $B_K$ as will be explained
shortly.

\section{QCD with chirally rotated mass terms}

Axial and vector symmetries can be distinguished according 
to whether or not the quark mass term is left invariant. 
Hence, a non-standard form of the quark mass term 
also modifies the form of the symmetry transformations.
Let us consider the continuum theory for a light quark doublet $\psi$
and the $s$ quark including a chirally twisted mass term,
\begin{equation}
{\cal L}_f=\overline{\psi}\left(D\kern-6pt\slash
                +m+i\mu_q\gamma_5\tau^3\right)\psi
               + \bar{s}\left(D\kern-6pt\slash +m_{\rm s}\right)s.
\end{equation}
After a chiral rotation of the doublet fields
\begin{equation}
 \psi'    =\exp\left(i \alpha\gamma_5\frac{\tau^3}{2}\right)\psi,\qquad
 \overline{\psi}' 
 =\overline{\psi}\exp\left(i \alpha\gamma_5\frac{\tau^3}{2}\right),
\end{equation}
and with $\alpha$ chosen such that $\tan\alpha = \mu_{\rm q}/m$,
the Lagrangian reads
\begin{equation}
 {\cal L}'_f=\overline{\psi}'\left(D\kern-6pt\slash +m'\right)\psi'
               + \bar{s}\left(D\kern-6pt\slash +m_{\rm s}\right)s,\qquad
	       m'= \sqrt{m^2+\mu_q^2}.
\end{equation}
The field rotation re-defines the symmetries and maps composite fields, e.g. 
\begin{equation}
  O_{\rm VV+AA}' = \cos(\alpha)O_{\rm VV+AA} - i\sin(\alpha) 
  O_{\rm VA+AV}
                 =  -i O_{\rm VA+AV}\quad (\alpha=\pi/2).
\end{equation}
In particular, the operators are mapped to each other at 
``maximal twist'' $\alpha=\pi/2 \Leftrightarrow m=0$, 
where the quark mass is determined entirely by the  
chirally twisted mass parameter $\mu_{\rm q}$.
Hence, by using Wilson quarks with a maximally twisted
light quark doublet and a standard $s$ quark, the complicated operator 
mixing problem can be by-passed~\cite{Frezzotti:2000nk}.
An additional benefit consists in the elimination of
unphysical zero modes by the twisted mass term.
This allows for numerical simulations to  get close
to the physical situation: light, mass degenerate $u,d$ quarks
and a heavier $s$ quark. However,
most  results in lattice QCD are
currently obtained in the limit where the kaon is made
out of mass-degenerate $d$ and $s$ quarks. In this way
a quenched artefact is avoided~\cite{Sharpe:1994dc}, but 
it implies that the zero mode problem is back for the
standard Wilson $s$ quark.
In order to cope with mass-degenerate quarks 
we also consider a different set-up, where
the r\^oles of $s$ and $u$ quarks are interchanged.
The chirally twisted doublet is now $\psi=(s,d)$
and after the chiral rotation one finds
\begin{equation}
  O_{\rm VV+AA}' = \cos(2\alpha)O_{\rm VV+AA} 
  - i\sin(2\alpha) O_{\rm VA+AV}
                 =  -i O_{\rm VA+AV}\quad (\alpha=\pi/4).
\end{equation}
Again the operators are mapped to each other provided
$\alpha=\pi/4 \Leftrightarrow m=\mu_{\rm q}$. For
obvious reasons we will refer to the two set-ups 
as $\pi/2$ and $\pi/4$ scenarios, respectively.
The latter has the advantage that both
$d$ and $s$ quark masses can be decreased simultaneously 
without encountering unphysical zero modes. 
Setting $\alpha$ to $\pi/2$ or $\pi/4$ requires some
parameter tuning using known results for 
finite renormalization  constants. In addition one needs the
multiplicative operator renormalization constant.
The renormalization problem has been solved
non-perturbatively in~\cite{Guagnelli:2005zc}, using a finite volume
scheme based on the Schr\"odinger functional
and recursive finite size techniques~\cite{Jansen:1995ck}.

\section{Numerical simulations}
The numerical simulations have been performed using the O($a$) improved Wilson
quark action and the Wilson's plaquette action for the gauge fields.
4-5 $\beta$-values have been chosen in the interval $[6.0-6.45]$,
corresponding to lattice spacings $a=0.05-0.1\,{\rm fm}$, if the
scale is set by $r_0=0.5\,{\rm fm}$.
The lattice volumes range form $16^3\times 48$ to $32^2\times 72$.
Quark masses are tuned to achieve $\alpha=\pi/2$ or $\alpha=\pi/4$
and pseudoscalar masses around or above $m_K$.
The analysis of excited states determined safe plateaux regions
where $B_K$ could be extracted from suitable ratios
of correlation functions.
Finite volume effects were checked at the coarsest lattice spacing 
($\beta=6.0$), and found to be below the statistical errors.
In the $\pi/2$ scenario chiral extrapolations were performed linearly
to the physical kaon mass. In the $\pi/4$ scenario 
the kaon mass could be reached by {\em interpolation}
except at $\beta=6.45$ where finite volume effects at the
physical kaon mass would have been non-negligible.
The resulting values for the $B_K$ are given in figure~1 as a function
of $a/r_0$. Discarding the values at the coarsest lattice the
data seem to scale very well. A continuum extrapolation of $\pi/4$
data linear in $a$ seems adequate.

%\psdraft
\begin{figure}
\centerline{
\psfig{file=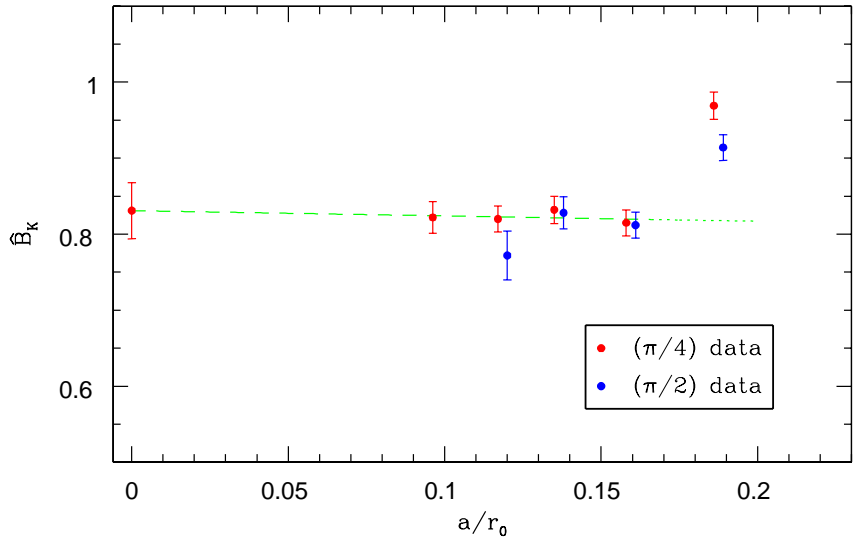,  
 width=14cm,bbllx=18,bblly=144,bburx=592,bbury=490}}
\vspace*{-1.5cm}
\caption{$B_K$ data for both $\pi/4$ and $\pi/2$ scenarios. The $\pi/2$
data is slightly shifted to the right to enhance readibility. Also shown
is the linear continuum extrapolation of $\pi/4$ data excluding the data on the
coarsest lattice.}
\end{figure}

\section{Conclusions}

We have performed a benchmark calculation of $B_K$ where,
apart from the quenched approximation, all systematic
errors are under control. These include cutoff effects,
finite volume effects and the contamination by excited states. 
A non-perturbative renormalization procedure has been employed, 
and the continuum limit has been taken.
The  preliminary result 
\begin{equation}
 \hat{B}_K = 0.834(37) \qquad \Leftrightarrow \qquad
  B_K^{\overline{\rm MS}}(2\,{\rm GeV})=0.604(27), 
\end{equation}
has the desired precision. Within errors it is 
compatible with results obtained by other groups using
different lattice regularisations
(see \cite{Dawson} for a recent review and further references). 
Hence, further progress will require 
the inclusion of dynamical quark flavours. The recent progress
with simulation algorithms for Wilson-type quarks~\cite{Luscher:2005mv} 
will soon allow to compute $B_K$ with a similar precision
including the effect of light dynamical up and down quarks.

\section*{Acknowledgments}

I would like to thank the organisers for the invitation
to this nice conference, my collaborators
P.~Dimopoulos, C.~Pena,~F. Palombi,
A.~Vladikas, J.~Heitger for a pleasant collaboration,
and the computer centre at DESY Zeuthen for providing 
the necessary CPU time on the APEmille machines.

%\section*{References}


\begin{thebibliography}{99}

%\cite{Battaglia:2003in}
\bibitem{Battaglia:2003in}
  M.~Battaglia {\it et al.},
  %``The CKM matrix and the unitarity triangle,''
  arXiv:hep-ph/0304132.
  %%CITATION = HEP-PH 0304132;%%

%\cite{Bernard:1987pr}
\bibitem{Bernard:1987pr}
  C.~W.~Bernard, T.~Draper, G.~Hockney and A.~Soni,
  %``Recent Developments In Weak Matrix Element Calculations,''
  Nucl.\ Phys.\ Proc.\ Suppl.\  {\bf 4} (1988) 483.
  %%CITATION = NUPHZ,4,483;%%

%\cite{Aoki:1999gw}
\bibitem{Aoki:1999gw}
  S.~Aoki {\it et al.}  [JLQCD Collaboration],
  %``The kaon B-parameter with the Wilson quark action using chiral Ward
  %identities,''
  Phys.\ Rev.\ D {\bf 60} (1999) 034511
  [arXiv:hep-lat/9901018].
  %%CITATION = HEP-LAT 9901018;%%

%\cite{Frezzotti:2000nk}
\bibitem{Frezzotti:2000nk}
  R.~Frezzotti, P.~A.~Grassi, S.~Sint and P.~Weisz  [Alpha collaboration],
%  ``Lattice QCD with a chirally twisted mass term,''
  JHEP {\bf 0108} (2001) 058 [arXiv:hep-lat/0101001].
%%CITATION = HEP-LAT 0101001;%%

%\cite{Sharpe:1994dc}
\bibitem{Sharpe:1994dc}
  S.~R.~Sharpe,
  %``Phenomenology from the lattice,''
  arXiv:hep-ph/9412243.
  %%CITATION = HEP-PH 9412243;%%


%\cite{Guagnelli:2005zc}
\bibitem{Guagnelli:2005zc}
  M.~Guagnelli, J.~Heitger, C.~Pena, S.~Sint and A.~Vladikas  [ALPHA
                  Collaboration],
%  ``Non-perturbative renormalization of left-left four-fermion operators in
%  quenched lattice QCD,''
  arXiv:hep-lat/0505002.
  %%CITATION = HEP-LAT 0505002;%%\end{thebibliography}

%\cite{Jansen:1995ck}
\bibitem{Jansen:1995ck}
  K.~Jansen {\it et al.},
%  ``Non-perturbative renormalization of lattice QCD at all scales,''
  Phys.\ Lett.\ B {\bf 372} (1996) 275
  [arXiv:hep-lat/9512009].
  %%CITATION = HEP-LAT 9512009;%%


\bibitem{Dawson}
C. Dawson, plenary talk at the XXIIIrd International Symposium on
Lattice Field Theory, 25-30 July 2005, Trinity College, Dublin, Ireland

%\cite{Luscher:2005mv}
\bibitem{Luscher:2005mv}
  M.~L\"uscher,
  %``Lattice QCD with light Wilson quarks,''
  arXiv:hep-lat/0509152.
  %%CITATION = HEP-LAT 0509152;%%

\end{thebibliography}
\end{document}